\documentstyle[12pt]{article}

\catcode`\@=11
\long\def\@makefntext#1{
\protect\noindent \hbox to 3.2pt {\hskip-.9pt
$^{{\ninerm\@thefnmark}}$\hfil}#1\hfill}                

 \def\@makefnmark{\hbox to 0pt{$^{\@thefnmark}$\hss}}  
 
\def\ps@myheadings{\let\@mkboth\@gobbletwo
\def\@oddhead{\hbox{}
\rightmark\hfil\ninerm\thepage}
\def\@oddfoot{}\def\@evenhead{\ninerm\thepage\hfil
\leftmark\hbox{}}\def\@evenfoot{}
\def\sectionmark##1{}\def\subsectionmark##1{}}
 
\newcounter{sectionc}\newcounter{subsectionc}\newcounter{subsubsectionc}
\renewcommand{\section}[1] {\vspace{0.6cm}\addtocounter{sectionc}{1}
\setcounter{subsectionc}{0}\setcounter{subsubsectionc}{0}\noindent
	{\bf\thesectionc. #1}\par\vspace{0.4cm}}
\renewcommand{\subsection}[1] {\vspace{0.6cm}\addtocounter{subsectionc}{1}
	\setcounter{subsubsectionc}{0}\noindent
	{\it\thesectionc.\thesubsectionc. #1}\par\vspace{0.4cm}}
\renewcommand{\subsubsection}[1] {\vspace{0.6cm}\addtocounter{subsubsectionc}{1}
	\noindent {\rm\thesectionc.\thesubsectionc.\thesubsubsectionc.
	#1}\par\vspace{0.4cm}}

\newcounter{appendixc}
\newcounter{subappendixc}[appendixc]
\newcounter{subsubappendixc}[subappendixc]

\renewcommand{\appendix}[1] {\vspace{0.6cm}
	\refstepcounter{appendixc}
	\setcounter{figure}{0}
	\setcounter{table}{0}
	\setcounter{equation}{0}
	\renewcommand{\thefigure}{\Alph{appendixc}.\arabic{figure}}
	\renewcommand{\thetable}{\Alph{appendixc}.\arabic{table}}
	\renewcommand{\theappendixc}{\Alph{appendixc}}
	\renewcommand{\theequation}{\Alph{appendixc}.\arabic{equation}}
	\noindent{\bf Appendix \theappendixc #1}\par\vspace{0.4cm}}

\renewenvironment{thebibliography}[1]
	{\begin{list}{\arabic{enumi}.}
	{\usecounter{enumi}\setlength{\parsep}{0pt}
\setlength{\leftmargin 1.25cm}{\rightmargin 0pt}
	 \setlength{\itemsep}{0pt} \settowidth
	{\labelwidth}{#1.}\sloppy}}{\end{list}}
 
\newcounter{itemlistc}
\newcounter{romanlistc}
\newcounter{alphlistc}
\newcounter{arabiclistc}

\newcommand{\fcaption}[1]{
	\refstepcounter{figure}
	\setbox\@tempboxa = \hbox{\tenrm Fig.~\thefigure. #1}
	\ifdim \wd\@tempboxa > 6in
	   {\begin{center}
	\parbox{6in}{\tenrm\baselineskip=12pt Fig.~\thefigure. #1}
	    \end{center}}
	\else
	     {\begin{center}
	     {\tenrm Fig.~\thefigure. #1}
	      \end{center}}
	\fi}
 
\newcommand{\tcaption}[1]{
	\refstepcounter{table}
	\setbox\@tempboxa = \hbox{\tenrm Table~\thetable. #1}
	\ifdim \wd\@tempboxa > 6in
	   {\begin{center}
	\parbox{6in}{\tenrm\baselineskip=12pt Table~\thetable. #1}
	    \end{center}}
	\else
	     {\begin{center}
	     {\tenrm Table~\thetable. #1}
	      \end{center}}
	\fi}
 
\def\@citex[#1]#2{\if@filesw\immediate\write\@auxout
	{\string\citation{#2}}\fi
\def\@citea{}\@cite{\@for\@citeb:=#2\do
	{\@citea\def\@citea{,}\@ifundefined
	{b@\@citeb}{{\bf ?}\@warning
	{Citation `\@citeb' on page \thepage \space undefined}}
	{\csname b@\@citeb\endcsname}}}{#1}}
 
\newif\if@cghi
\def\cite{\@cghitrue\@ifnextchar [{\@tempswatrue
	\@citex}{\@tempswafalse\@citex[]}}
\def\citelow{\@cghifalse\@ifnextchar [{\@tempswatrue
	\@citex}{\@tempswafalse\@citex[]}}
\def\@cite#1#2{{$\null^{#1}$\if@tempswa\typeout
	{IJCGA warning: optional citation argument
	ignored: `#2'} \fi}}

\def\fnt#1#2{\footnotetext{\kern-.3em
	{$^{\mbox{\sevenrm #1}}$}{#2}}}
 
 1
 1
 1

\font\tenrm=cmr10

\font\ninerm=cmr9

\begin{document}

\begin{Large}
\begin{center}
{\bf SPACETIME  FOAM  AND THE ELECTROWEAK COUPLING CONSTANT}
\end{center}
\end{Large}

\vspace{1 cm}

\vspace{5 mm}
\hspace{2 cm}  J.L. ROSALES \footnote{E-mail: rosales@besaya.unican.es}

\hspace{2 cm}
     {\em Departamento de F\'isica Moderna, Facultad de Ciencias }

\hspace{2 cm} {\em Universidad de Cantabria, 39005, Santander.}

\vspace{5 mm}
\begin{quote}
\begin{small}
We compute  the regularized temperature for a spacetime foam
model, consisting on $\mbox{$\cal{S}$}^4$ instantons, in quantum gravity.
Assuming that thermal equilibrium takes place with some amount of  
radiation - with thermal
fields in the {\bf SU}(2)x{\bf U}(1) gauge theory - we obtain
the remarkable result that the squared value of this temperature 
exactly coincides  with 
the electroweak coupling constant at the energy scale
of the gauge bosons $W^{\pm}$. This is consistent with the 
classical ADM result that the  electrical charge 
should be equal to its {\it finite} gravitational self energy.

\end{small}
\end{quote}

\vspace{2 mm}

{\it Introduction.}

 Wheeler\cite{kn:Wheeler} has been the first to point out that
quantum fluctuations of spacetime are inescapable if   
we believe the quantum uncertainty principle and Einstein's theory of
general relativity for the graviational field, thus, at a submicroscopic 
scale the geometry and its corresponding topology {\it would resonate} between
one configuration and another.
This  gives spacetime a foamlike structure though it
looks smooth on large scales compared to the Planck scale. 

Some years later Hawking has developped some further ideas based on his experience
concerning the properties of quantum black holes\cite{kn:bhlaw}-\cite{kn:Hawking0}. 
He supposed
that Wheeler's foam would eventually consist on a see of virtual black holes
or gravitational instantons\cite{kn:Eguchi}. He refered to that the
{\it quantum bubbles picture}. The  idea comes originated from making an
analogy between the Ernst solution of Einstein-Maxwell equations\cite{kn:Ernst}
- representing
two charged black holes accelerating away from each other in a spacetime
that is asymptotically the Melvin universe- and the pair creation of ordinary
charged particles. An electron and a positron emerge from tunneling through
Euclidean space as a pair of real particles in Minkowski space. The analogy
with the Ernst solution -whose Euclidean topology is  
$\mbox{$\cal{S}$}^2 \mbox{x}\mbox{$\cal{S}$}^2$ minus a point sent to infinity-
indicates that the typical quantum bubble is just the topological sum of the  
compact bubble $\mbox{$\cal{S}$}^2 \mbox{x}\mbox{$\cal{S}$}^2$ with the 
non compact space $\mbox{$\cal{R}$}^4$. These black holes need not to carry
electric charge and are not in general solutions of Einstein's equations but
they would occur as quantum fluctuations.

Moreover, if black holes are present, the 
particles inmersed in the spacetime foam should
not be transparent to the foamy structure in such a way that the constants 
of nature (as for instance the electrical charges) should, 
consequently, be determined from their interaction 
within the foam. Other observational consequences would be the loss of
quantum coherence but this effect depends strongly on the spin of the field.
The effective interactions induced by bubbles, upon computing the scattering 
matrix of the fields in the  virtual black hole metric, are suppressed by factors
of the Planck mass. The only exception are scalar fields. This means
that we will never observe the Higgs particle\cite{kn:bubbles}.

Among other most direct consequences of quantum gravity is the existence of
a topological entropy. This implies that microscopically we can only
make statistical descriptions and, the main interation is governed in
terms of an averaged temperature. This feature could be  a consequence 
of the statistical equilibrium between topological configurations in the
spacetime foam. Thus, Bousso\cite{kn:Bousso} has made the hypothesis that the topology
of the $\mbox{$\cal{S}$}^2 \mbox{x}\mbox{$\cal{S}$}^2$ bubble 
spontaneously changes to $\mbox{$\cal{S}$}^4$ which is
conformally equivalent to flat Euclidean space $\mbox{$\cal{R}$}^4$ plus
a point added at infinity.
These speculations allow us to consider  simpler models for the
quantized spacetime.

Consider that the Euclidean metric is given by that of the  $\mbox{$\cal{S}$}^4$,
\begin{equation}
ds^2= d\tau^2+a^2 \cos^2(\tau/a) d\Omega_3^2
\end{equation}
then the action is that of an Einstein space (such that 
$R_{\mu\nu}=\Lambda g_{\mu\nu}$ ) with positive cosmological
constant $\Lambda= 3/a^2$, i.e.,
\begin{equation}
I=-\frac{3 \pi}{2\Lambda} \mbox{,}
\end{equation}
whose entropy is simply
\begin{equation}
S=\pi a^2 \mbox{.}
\end{equation}
Notice that the metric of the instanton is periodic in the Euclidean time
direction with a period $\beta= 2\pi a$. This means that Green functions
are also periodic with that period and behave as partition functions of
a thermal ensamble with a temperature $T=\beta^{-1}$.  Moreover, a 
suitable thermal energy can be obtained quite straightfowardly 
\begin{equation}
u=\int\beta^{-1} dS =a \mbox{.}
\end{equation}
Yet, owing to the existence of some non-vanishing temperature we could 
formally consider  the black body radiation inside a given three 
dimensional cavity. This radiation would heuristically correspond 
to the spontaneous polarization of vacuum near the horizon of the virtual black hole
metric.

Boltzmann's equilibrium takes place for
a configuration that maximizes the entropy for a  value of the total
energy.
\begin{equation}
S_T=\pi a^2+\frac{4}{3}\sigma V T^3 \mbox{,}
\end{equation}
\begin{equation}
E=a+\sigma VT^4 \mbox{,}
\end{equation}
where, $\sigma=N \pi^2/30 $ is Stefan's constant for the total number of 
thermal species. Now,
by eliminating  $T$ and defining $\omega=4/3(\sigma V)^{1/4}$.
\begin{equation}
T=\frac{4}{3}\omega^{-1}(E-a)^{1/4} \mbox{.}
\end{equation}

Yet, the maximal total entropy configuration is obtained for $\omega$
satisfying the constraint $\partial_a S_T =0$, or
\begin{equation}
\omega=\frac{8\pi}{3}a (E-a)^{1/4} \mbox{,}
\end{equation}
and, from (7) and (8), the equilibrium temperature is that of the instanton
\begin{equation}
T^{-1}=2\pi a \mbox{.}
\end{equation}
Equation (9) states  the stability of the horizon under generic  
perturbations in the temperature.

Now, let us obtain that portion of the instanton energy, $E_0$, corresponding to
the quantum fields in statistical equilibrium having this intrinsic gravitational 
thermal energy 
(i.e., the energy corresponding to radiation). 

Equation (5) can be written as
\begin{equation}
S=\pi a^2 +\omega (E-a)^{3/4} \mbox{.}
\end{equation}
Defining
$\Omega=S/\pi E^2$, $\varepsilon= a/E$
and  $\omega\rightarrow \omega E^{-1/4} /\pi$,
we obtain 
\begin{equation}
\Omega=\varepsilon^2+\omega (1-\varepsilon)^{3/4} \mbox{.}
\end{equation}
This function was discovered by the first time 
by Gibbons and Perry related with the
problem of the condensation of a black hole from pure radiation in a box\cite{kn:Bhcondens}.
Its absolute maximun is obtained for  $\varepsilon_0\simeq 0.977015 $   
and $\omega\simeq 1.014$ \footnote {see Appendix 1};
from them we compute  the portion of the energy  corresponding to radiation 
\begin{equation}
E_{0}= \frac{E_{rad}}{\gamma} \mbox{,}
\end{equation}
the proportional constant being $\gamma=1/\varepsilon_0 - 1= 0.02352482346\cdots$;
this turns out to be a very strong conditition for the physics of
the instanton boundary.

Following this thermodynamical constraint, 
we get for the total gravitational energy density
\begin{equation}
\rho=\frac{\rho_{rad}}{\gamma}=\frac{\Lambda}{8\pi}+\rho_{rad}+\rho_{reg} \mbox{\hspace{2 mm},}
\end{equation}
here, $\rho_{rad}$ and $\rho_{reg}$ are, respectively, 
the thermal energy density and some
regularized energy density coming from substracting out the infinities 
of the  zero point energy in field theory.
The regularized energy density is \footnote {see Appendix 2 }
\begin{equation}
\rho_{reg}=\frac{C}{480\pi^2 a^4}=\frac{C\pi^2}{30}T^4 \mbox{\hspace{2 mm},}
\end{equation}
where $C$ is the number of spins. 

Yet, recalling that $\rho_{rad}=\sigma T^4$ and 
$\Lambda=12 \pi^2 T^2$,
we finally obtain
\begin{equation}
\frac{3\pi}{2}T^2=\frac{\pi^2 N}{30} T^4\{\frac{1}{\gamma}-1- \frac{C}{N}\} \mbox{\hspace{2 mm},}
\end{equation}
i.e., if $\gamma_{*}^{-1}=\gamma^{-1}-1-C/N$ is the regularized value of $\gamma$,
\begin{equation}
T^2 (N,C)=\frac{\gamma_{*}(N,C)}{N}\frac{45}{\pi} \mbox{ \hspace{2 mm}.}
\end{equation}

This is the thermodynamical constraint. It generates, consequently, 
the possible values of the gravitationally renormalized 
mass. 

\vspace{6 mm}

{\it Gravitational self energy of charged particles}.

Classical theory predicts that the total mass of a charged particle
would arise from its coupling to the field. 
Moreover, after the results of Arnowitt, Deser and 
Misner\cite{kn:ADM}, it has 
been rigurously demostrated that general relativity predicts a finite
value of the total gravitational  self energy of a classical electron, 
independent of its mechanical mass and completly determined by its
charge. This can be understood on heuristical grounds from the fact
that general relativity effectively replaces the  mechanical  mass $m_0$ by 
by $m$, the total self energy, in the interaction term:
$m=m_0-\frac{1}{2} m^2/r +\frac{1}{2}e^2/r$.
$m$ is therefore determined by the 
classical equation  
\begin{equation}
m=-r +[r^2 +e^2 +2m_0 r ]^{1/2} \mbox{.} 
\end{equation}
Yet, while for $r\gg 1$ the Newtonian limit is recovered,
\begin{equation}
m\sim m_0 - m_0^2/2r +e^2/2r +\cdots \mbox{,}
\end{equation}
in the limit $r\rightarrow 0$ (point like particles), we have instead the finite result
$m=e\equiv \alpha^{1/2}$, independently of the mechanical mass $m_0$.
This figure is of the order of the Planck mass for the electron charge.

On the other hand, the available energy of 
charged particles inmersed within the spacetime foam 
should be the intrinsic  temperature (a quantity also of the order 
of Planck mass) and, we may be curious about the possibility that
it were to coincide  with the
value of its own  classically predicted self energy, 
i.e., the electric charge at a given energy scale.

This hypothesis would be  tested  in the following section
upon computing exactly the value of the foamy temperature in our model 
(for a radiation of thermalized  fields). 
The only additional imput we would require is
the spin statistics of the  fields
in the standard model of partile physics.

\vspace{6 mm}

{\it The fine structure constant}.

The fact that $\alpha^{1/2}$ could be the actual gravitational zero point energy, $T$, 
motivates adapting the thermodynamical constraint in Eq. (16)  
for the  theory of  electroweak interactions. 
From this field theory in three dimensions one obtains straightforwarly,
by counting  the total number of available fermionic
and bosonic states,
$N_{1}=191/4$ and $C_1=53$ (we have discarded the Higgs boson spin state
since, following the quantum bubbles picture, the quantum coherence of the
scalar field should be lost\cite{kn:bubbles}\footnote {Appendix 3}.)  
On the other hand, since there is no external electromagnetic field, we could select
the gauge so that $A_{\mu}$ be identically zero. This requires, in
the {\bf SU}(2)x{\bf U}(1) gauge theory, the generation of a pair
of radiative $W^{\pm}_{\mu}$ vector bosons at that energy scale,
\begin{equation}
A'_{\mu}(x)=A_{\mu}(x)+(1+(\frac{2g_1}{g})^2)^{1/2}\partial_{\mu} \varepsilon (x)= 0 \nonumber
\end{equation}
\begin{equation}
W'^{\pm}_{\mu}(x)=W^{\pm}_{\mu}(x)[1 \pm 2i g_1\varepsilon(x)] \mbox{\hspace{2 mm},}
\end{equation}
here $g_1$ and $\varepsilon (x)$  are the parameters of the {\bf U}(1) gauge
symetry   and $g$ is the gauge charge of the {\bf SU}(2) group.

In this latter case, the number of thermal species is $N_{2}=N_1-2=183/4$ and $C_2=C_1-2=51$. 
From the same thermodynamical constraint (16) we get
\begin{equation}
\alpha_{W}= T^2(183/4,51) =\{\frac{\gamma}{1-\gamma(1+68/61)}\}\frac{180}{183 \pi} \simeq 129.01^{-1}\mbox{,}
\end{equation}
a figure that corresponds (almost exactly) to the radiative electroweak coupling 
constant to that scale. Recall that 
$\alpha_{Z^0}^{-1}=128.878\pm 0.090$ - see \cite{kn:Weinberg}\cite{kn:cern}-  at the
sliding scale of the $M_W$, we have
\begin{equation}
\alpha_{W}=\frac{\alpha_{Z^{0}}}{1-\frac{2}{3 \pi}\alpha_{Z^{0}}
[\log[\frac{M_{W}}{M_{Z^{0}}}]-\frac{5}{6}]}= [129.08\pm 0.09] ^{-1}
\end{equation}
This  seems to confirm the quantum bubbles picture
predictions of Hawking (since counting the Higgs state would have lead to a quite 
less exact figure). Moreover we have to remark that (21) only corresponds to the
charge of the intermediate bosons $W^{\pm}$ and that we 
have not been capable of predicting quark charges for 
it is not possible to  isolate the  quarks at arbitrary large distances
where the ADM mass is defined.

In summary, what we have calculated in this paper is the actual value of the 
electric charge that is compatible with a foamlike structure of virtual black holes.
This means that if it does happen that a charge is present, then the value of
this charge becomes generated as a consequence  of 
the gravitational field vacuum.

\vspace{4 mm}

\noindent{\it Acknowledments}. 
I want to express my   
warmest thankfulness to the Department of Theoretical Physics of the Universidad
Aut\'onoma de Madrid for kind hospitality during the completion of this
work. This work is done after a contract of the Spanish Ministry of Education and
Culture under the project C.I.C. y T. PB95-594.

\newpage

{\bf Appendix 1}

The maximum value of  $\Omega$ in (11) satisfies 
$\partial_{\varepsilon} \Omega=0$, i.e.,
\begin{eqnarray*}
\omega=\frac{8\varepsilon}{3}(1-\varepsilon)^{1/4} \mbox{, \hspace{62 mm} (1.1 }  
\end{eqnarray*}
On the other hand, the absolute maximun of $\Omega$ is still at $\varepsilon=0$
unless there were  some $\varepsilon_{0}$ that
\begin{eqnarray*}
\Omega(\varepsilon_0)=\Omega(0)=\omega \mbox{,\hspace{60 mm} (1.2}
\end{eqnarray*}
this implies 
\begin{eqnarray*}
\omega=\frac{\varepsilon_0^2}{1-(1-\varepsilon_0)^{3/4}} \mbox{,\hspace{56 mm} (1.3}
\end{eqnarray*}
Using  (1.1) and (1.2) we obtain 
\begin{eqnarray*}
1-\varepsilon_0 =(1-\frac{5\varepsilon_0}{8})^4  \mbox{,\hspace{57 mm} (1.4}
\end{eqnarray*}
From the solution of this algebraic equation we get 
\begin{eqnarray*}
\gamma=\frac{1}{\varepsilon_0}-1=0.023524823\cdots \mbox{.\hspace{38 mm} (1.5}
\end{eqnarray*}

\vspace{6 mm}

{\bf Appendix 2}

Vacuum energy is a devergent quantity in field theory, for instance, in 
Minkowski space
\begin{eqnarray*}
<0|H0>=<0|0>\sum_{\vec{k},\sigma}\frac{1}{2}\omega_{\vec{k},\sigma} \mbox{,\hspace{41 mm} (2.1}
\end{eqnarray*}
where $\sigma$ is the polarization (or spin) degree of freedom. 

The analogous to this quantity in curved space is
\begin{eqnarray*}
E=\frac{1}{2}\int d\mu (k,\sigma) \frac{k}{a} \mbox{,\hspace{57 mm} (2.2}
\end{eqnarray*}
where we introduce the normal modes labelled by $k$ and $a$ is the radius of
the $S^3$. We have, neglecting mass contributions, 
\begin{eqnarray*}
E=\frac{1}{2}\sum_{\sigma}\sum_{k, J,M} \frac{k}{a} 
\end{eqnarray*}
\begin{eqnarray*}
E=C\frac{1}{2}\sum_{k=1}^{\infty}k^2\frac{k}{a}
\end{eqnarray*}
\begin{eqnarray*}
=\lim_{s\rightarrow -1} 
C\sum_{k=1}^{\infty}k^2(\frac{k}{2a})^{-s}
\mbox{,\hspace{50 mm} (2.3}
\end{eqnarray*}
where, $J=0,\cdots, k-1$ and $M=-J,\cdots, +J$ are the labels of the spherical harmonics
and $C=\sum_{\mbox{\begin{small} fields \end{small}}}(2s_i+1)$ is the total number
of spins.

The sum gives a finite result in the complex plane, obtaining
\begin{eqnarray*}
E=C\frac{\zeta(-3)}{2a}= \frac{C}{240 a} \mbox{.\hspace{54 mm} (2.4}
\end{eqnarray*}
Dividing E by the proper volume of space $V=2\pi^2 a^3$, the  energy density is
simply
\begin{eqnarray*}
\rho=\frac{C}{480 \pi^2 a^4} \mbox{.\hspace{70 mm} (2.5}
\end{eqnarray*}

\vspace{6 mm}

{\bf Appendix 3}

The number of thermal species,$N$, and the number of spins, $C$, in the Weinberg -Salam  model is 
calculated straightforwardly as follows:

The corresponding number of spin states for a model with
three  quark generations (i.e., six quarks and six antiquarks)  is
\begin{eqnarray*}
N_{q}=12\cdot  2 \cdot 7/8 \mbox{,\hspace{6 mm}} C_q=24
\end{eqnarray*}
where we  took into account  a factor  two
for spin and the typical $7/8$ for 
Fermi Statistics.

For charged leptons we get, analogously
\begin{eqnarray*}
N_{CL}= 6\cdot 2\cdot 7/8\mbox{.\hspace{6 mm}} C_{CL}=12
\end{eqnarray*}

For neutrinos 
\begin{eqnarray*}
N_{\nu}=3\cdot  2 \cdot 7/8 \mbox{,\hspace{6 mm}} C_{\nu}=6
\end{eqnarray*}
where we have considered the same particle  for neutrinos and  antineutrinos.

For vector bosons
\begin{eqnarray*}
N_{W,Z^{0}}= 3 \cdot 3 \cdot 1 \mbox{,\hspace{6 mm}} C_{W,Z^{0}}=9
\end{eqnarray*}
and for the electromagnetic field,
\begin{eqnarray*}
N_{A^{\mu}}=2 \mbox{,\hspace{6 mm}} C_{A^{\mu}}=2
\end{eqnarray*}
here we have replaced the previous $7/8$ factor by $1$ in Bose Statistics. 

The total number of thermal species is
\begin{eqnarray*}
N_{1}=N_q+N_{CL}+N_{\nu}+N_{W,Z^{0}}+N_{A^{\mu}}= 191/4 \mbox{,\hspace{6 mm}} C_1=53
\end{eqnarray*}

Gauging out the electromacnetic field would be equivalent to obtaining
a radiative pair of charged bosons. This leads to the number
\begin{eqnarray*}
N_{2}=N_1-2=183/4 \mbox{,\hspace{6 mm}} C_2=C_1-2=51 \mbox{.}
\end{eqnarray*}
\vspace{ 4 mm}


\vspace{6 mm}

\begin{thebibliography}{99}


\bibitem{kn:Wheeler}
J. A. Wheeler, {\em Ann. Phys.} (NY),{\bf 2}, 604 (1957).

\bibitem{kn:bhlaw}
S. W. Hawking, {\em Commun. Math. Phys.}, {\bf 43}, 99 (1975).

\bibitem{kn:Hawking0}
S. W. Hawking, {\em Nucl. Phys.}, {\bf 144}, 349 (1978).

\bibitem{kn:Eguchi}
T. Eguchi, P. B. Gilkey and, A. J. Hanson, {\em Phys. Rep.}, {\bf 66}, 213 (1980).

\bibitem{kn:Ernst}
F. J. Ernst, {\em J. Math. Phys.}, {\bf 17}, 515 (1976).

\bibitem{kn:bubbles}
S. W. Hawking, {\em Phys. Rev.}, {\bf 53}, 3099 (1996).

\bibitem{kn:Bousso}
R. Bousso, {\em Phys. Rev.} D{\bf 58}, 083511 (1998).

\bibitem{kn:Bhcondens}
G. W. Gibbons and M. J. Perry, {\em Proc. R. Soc. London}, A{\bf 358}, 467 (1978).

\bibitem{kn:ADM}
R. Arnowitt, S. Deser, and C. W. Misner, "{\it The dynamics
of general relativity}" in {\it Gravitation: An Introduction
to Current Research}" pp. 227 - 265, L. Witten, ed; Wiley, New York (1962).

\bibitem{kn:Weinberg}
S. Weinberg, "{\it The Quantum Theory of fields II}, Cambridge University Press 1996.

\bibitem{kn:cern}
"{\em  European organization for particle physics:
A Combination of Preliminary Electroweak Measurements and Constraints on the
Standard Model}", Preprint CERN-EP/99-15.


\end{thebibliography}
\end{document}